\documentclass[preprint,12pt,authoryear]{elsarticle}

\usepackage{amssymb}

\usepackage[margin=2cm]{geometry}

\usepackage{balance}

\usepackage{wrapfig}
\usepackage{comment}
\usepackage{array}
\usepackage{hyperref}
\usepackage{tablefootnote}
\usepackage{amsmath}
\usepackage{cleveref}

\usepackage{xcolor}

\hypersetup{
	colorlinks,
	linkcolor={red!50!black},
	citecolor={blue!70!black},
	urlcolor={blue}
}
\usepackage{multirow}
\usepackage{epstopdf}
\usepackage{booktabs}
\usepackage{subcaption}
\usepackage{caption}
\usepackage{enumerate}
\usepackage{adjustbox}

\usepackage{eurosym}
\usepackage{graphicx,multicol}
\usepackage{xspace}
\usepackage{soul} 
\usepackage{float}

\usepackage[normalem]{ulem}

\newcommand\eg{\emph{e.g.,}\xspace}
\newcommand\ie{\emph{i.e.,}\xspace}
\newcommand\etal{\emph{et al.}\xspace}
\newcommand\COVID{{COVID-19}\xspace}

\graphicspath{{Figures>}{<Figures/Facebook}{<Figures/OpenRTB>}}

\journal{Telematics and Informatics}

\begin{document}

\begin{frontmatter}



\title{How resilient is the Open Web to the COVID-19 pandemic?}


\author[inst1]{José González-Cabañas\corref{cor1}}
\author[inst2]{Patricia Callejo}
\author[inst3]{Pelayo Vallina}
\author[inst1,inst2]{\\Ángel Cuevas}
\author[inst1,inst2]{Rubén Cuevas}
\author[inst3]{Antonio Fernández Anta}

\affiliation[inst1]{organization={Department of Telematic Engineering, Universidad Carlos III de Madrid},
        city={Leganés},
        postcode={28911},
        country={Spain}}

\affiliation[inst2]{organization={UC3M-Santander Big Data Institute, Universidad Carlos III de Madrid},
        city={Getafe},
        postcode={28903},
        country={Spain}}

\affiliation[inst3]{organization={IMDEA Networks Institute},
        city={Leganés},
        postcode={28911},
        country={Spain}}
        
\cortext[cor1]{Corresponding author: jgcabana@it.uc3m.es}


\begin{abstract}
In this paper we refer to the Open Web to the set of services offered freely to Internet users, representing a pillar of modern societies. Despite its importance for society, it is unknown how the COVID-19 pandemic is affecting the Open Web.
In this paper, we address this issue, focusing our analysis on Spain, one of the countries which have been most impacted by the pandemic.  

On the one hand, we study the impact of the pandemic in the financial backbone of the Open Web, the online advertising business. To this end, we leverage concepts from Supply-Demand economic theory to perform a careful analysis of the elasticity in the supply of ad-spaces to the financial shortage of the online advertising business and its subsequent reduction in ad spaces' price. On the other hand, we analyze the distribution of the Open Web composition across business categories and its evolution during the COVID-19 pandemic. These analyses are conducted between Jan 1st and Dec 31st, 2020, using a reference dataset comprising information from more than 18 billion ad spaces.

Our results indicate that the Open Web has experienced a moderate shift in its composition across business categories. However, this change is not produced by the financial shortage of the online advertising business, because as our analysis shows, the Open Web's supply of ad spaces is inelastic (i.e., insensitive) to the sustained low-price of ad spaces during the pandemic. Instead, existing evidence suggests that the reported shift in the Open Web composition is likely due to the change in the users' online behavior (e.g., browsing and mobile apps utilization patterns).
\end{abstract}




\end{frontmatter}


\section{Introduction}
\label{sec:introduction}

The COVID-19 pandemic has affected almost every single corner of modern society. 
The research community is tirelessly trying to understand the impact of the COVID-19 pandemic in different aspects and sectors so that our society is better prepared to face the rest of this pandemic, as well as future ones. 
For instance, the computer science community has contributed literature on how to fight the pandemic~\citep{habes2020relation}, on changes on the citizens’ mobility patterns \citep{lutu2020characterization,santamaria2020measuring,schlosser2020covid}, and on the resilience of fixed and mobile networking infrastructures~\citep{internetReactedCovidFB,candela2020impact,feldmann2020lockdown}.
In this paper, we contribute to this effort by evaluating the impact that the COVID-19 pandemic has had in the Open Web. Note that we define the term \emph{``Open Web''} define as the collection of Internet services that users can access for free (comprising the most popular online services such as web pages, mobile apps, free video platforms, or social media platforms). To make the context of this paper clearer, each time we talk about Open Web we refer to it as this definition. We know there could be other definitions for the term Open Web. The Open Web is one of the main pillars of developed societies nowadays, and therefore any impact that the pandemic has on it may have subsequent consequences in the society.

It is commonly agreed that the COVID-19 pandemic has, in general, impacted the citizens' behavior due to different limitations imposed on mobility, the massive movement to teleworking,  etc. Subsequently, the enforced limitation on the regular citizens' activity, has impacted economically and financially a large number of businesses. Following this, we will focus our analysis of the impact of the COVID-19 pandemic in the Open Web on two aspects which may have triggered the transformation of the Open Web: (1) the financial impact and (2) the impact of changes in citizens' online behavior.

\paragraph{Financial impact}
The online advertising business is the fundamental financial source of the Open Web. The major part of services in the Open Web obtain their revenue by providing ad spaces, which are later filled by advertisers in exchange for a fee.  Industry reports reveal that the COVID-19 pandemic has reduced the advertisers' investment in digital marketing, which has negatively affected the online advertising business \citep{iab:sellside:COVID, pixalate:programmatic:COVID,cnbc:fb:down}. This translates into a significant reduction in the ad spaces' demand \citep{iab:buyside:COVID, statista:impact:investment}. As the Supply-Demand economic theory states, and our empirical analysis proves, the shrink in demand led to a drop in ad spaces' price.

The first contribution of this paper consists of analyzing the resilience of the Open Web to the reported financial shortage produced by the COVID-19 pandemic in the online advertising business. To this end, we study the elasticity of the supply of ad spaces, by the Open Web, to the reported drop in ad spaces price and demand. We envision two possible scenarios: 
(1) If the Open Web offers an inelastic
supply (\ie the supply of ad spaces is not sensitive
to the drop in price and demand), we can conclude
that it is resilient to the reduction of income produced by the financial shortage.
Contrarily, (2) if the Open Web presents an elastic supply (\ie the supply is
sensitive to the variability in price and demand), we would
expect a reduction in the number of ad spaces offered by
the Open Web, which in turn would represent a reduction
in the number of services forming the Open Web. For instance,
some players may have opted for new monetization
schemes (\eg subscription models) or, even worse, some
players may have stopped their operation due to the lack of
financial sustainability. This means that the Open Web may
have shrunk, and might be at risk. 

\paragraph{Impact due to the change in users' online behavior}
Although there isn't (to the best of the authors' knowledge) a specific academic work looking into changes of users' online behavior caused by the pandemic, different non-academic reports \citep{similar:web:change, web:index:change} as well as academic papers showing a significant modification of the Internet traffic pattern \citep{feldmann2020lockdown, internetReactedCovidFB, candela2020impact} suggests the existence of such change. If this hypothesis is correct, the change in users’ behavior may have impacted the composition of the Open Web, \ie the relevance of different services may have changed. For instance, users may have reduced their visits to vacations and travel-related websites, but have increased their activity on gaming websites.
To study this specific aspect, we have analyzed the distribution into business categories of thousands of webpages and mobile apps that offered ad spaces across the different phases of the pandemic period. Using the Ružička index \citep{ruzicka18anwendung} we objectively measure if (and how much) the composition of the Open Web across business categories has changed with the pandemic.

In this paper, we have run the described analyses for Spain, one of the countries which has been most severely impacted by the \COVID pandemic in the number of cases and casualties. This led to severe mobility restriction, including a 2-month strict lockdown, as well as an important contraction of the economy (GDP shrank by 21.5\% in the first half of  2020 \citep{gdp:spain}). To conduct our empirical study, we leverage a dataset including data from more than 18.2B ads gathered from the bid request stream of the Sonata DSP~\citep{sonata} operated by TAPTAP Digital~\citep{taptap}, an online advertising stakeholder with a strong presence in the Spanish market. Moreover, we use a separate dataset to specifically analyze the impact of the pandemic on Facebook, which is a representative of a selected group of services with a dominant position in the Open Web and the online advertising ecosystem.


In summary, this paper provides the following novel contributions: (1) we formulate the relation between online advertising supply and the resilience of the Open Web to a crisis, like the \COVID pandemic; (2) we do a pioneering analysis of the elasticity of supply to price changes in online advertising; (3) we evaluate the changes in the composition of business categories of the Open Web by analyzing the changes in the supply of ads during the \COVID pandemic; (4) we exploit unique datasets characterizing the evolution of supply and price in online advertising.
\section{Background}
\label{sec:background}

In this section, we briefly describe the two online advertising markets covered in the paper: the Programmatic Advertising Market and the Facebook (FB) advertising platform. Finally, we describe the Supply-Demand theory in the context of online advertising.

\subsection{Programmatic Advertising Market}
\label{subsec:backgroud_programmatic}

The \emph{Programmatic Advertising Market} is the most important component of online advertising. For instance, in 2018 it generated 72,1\% (\euro16.8B) of the overall revenue of online advertising in Europe \citep{openrtb:iab:revenue}. This market distributes ads mainly through web pages and mobile apps, as part of the Open Web ecosystem.

\begin{figure*}[t!]
      \centering
      \begin{minipage}{0.55\linewidth}
          \begin{figure}[H]
            \includegraphics[width=\linewidth]{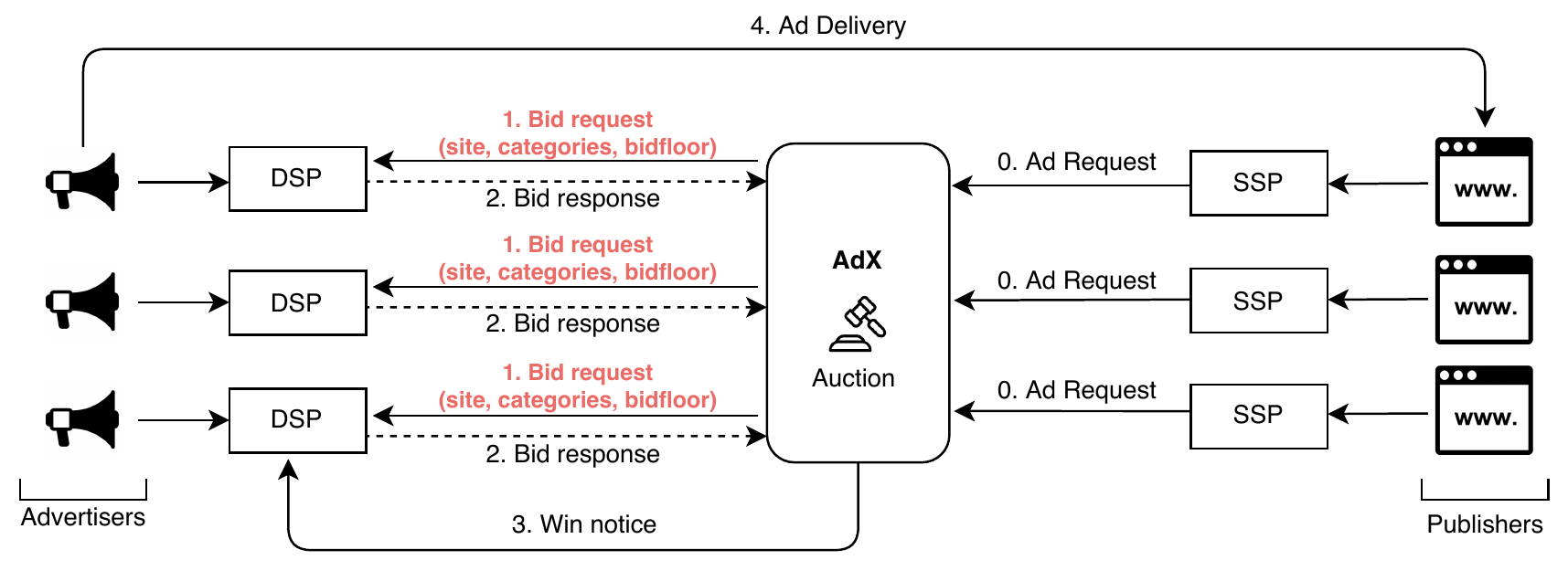}
             \caption{Programmatic Advertising Market operation.}
        \label{fig:programmatic_operation}
          \end{figure}
      \end{minipage} \quad
      \begin{minipage}{0.41\linewidth}
          \begin{figure}[H]
          \centering
          \includegraphics[width=0.9\linewidth]{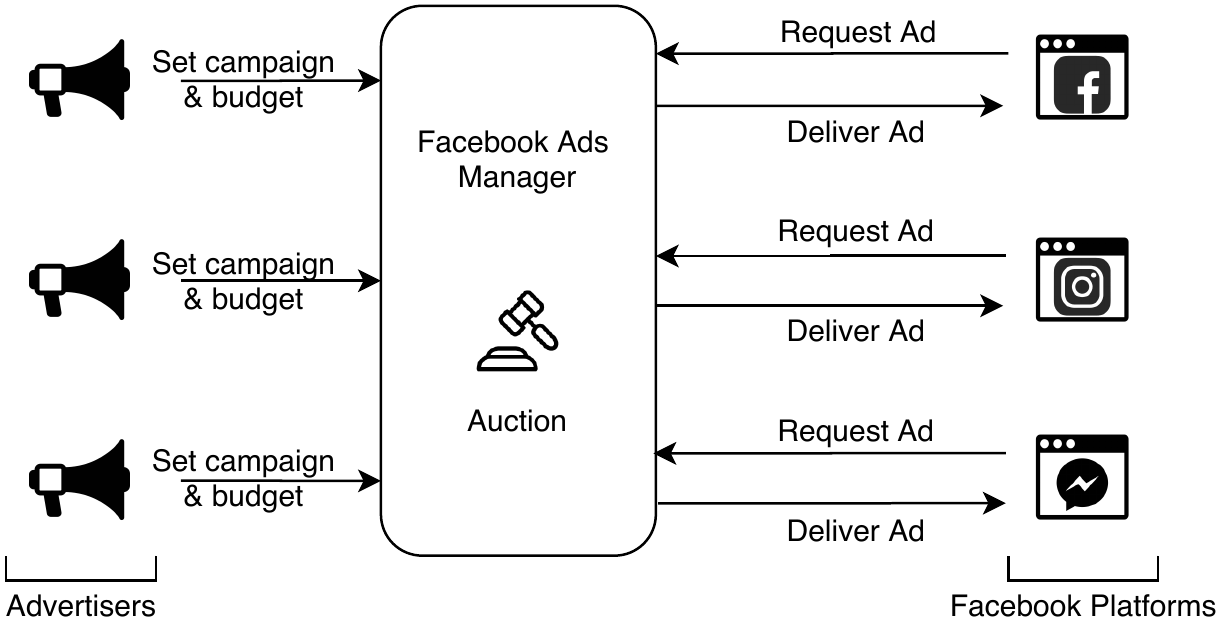}
              \caption{FB Advertising Platform operation.}
            \label{fig:fb_operation}
          \end{figure}
      \end{minipage}
\end{figure*}

\Cref{fig:programmatic_operation} describes the operation of the Programmatic Advertising Market. It is divided into three parts: \emph{sell-side}, \emph{buy-side}, and \emph{Ad Exchanges}.  
The sell-side (or supply-side) is composed of publishers (who own websites and mobile apps) offering ad spaces, and by ad networks or Supply Side Platforms (SSPs) that aggregate ad spaces from several publishers.
The buy-side (or demand side) is formed by advertisers willing to buy ad spaces. Advertisers (or media agencies on their behalf) buy the ad inventory from the sell-side through their online advertising campaigns, set up in Demand Side Platforms (DSPs). 
Finally, Ad Exchanges (AdXs) connect the buy and sell sides. For each ad space provided by an SSP (or ad network), the AdX runs an auction among its associated DSPs using the OpenRTB protocol \citep{openrtb}. In particular, the AdX sends to the DSPs a \emph{bid request} message including information related to the specific ad space (domain information, device information, and end-user information). A DSP checks if the ad space meets the requirements of any of its configured ad campaigns. If so, it replies with a bid response including the bidding price (the price the advertiser is willing to pay for that ad space). The AdX chooses the winning bid, notifies the winning DSP with a \emph{winning notice} message and its associated advertiser delivers its ad into the ad space. There are two parameters included in the bid request by AdXs relevant for this study:

\begin{itemize} 
\item[-] The \emph{bid floor}, the minimum allowed bidding price, is expressed in terms of CPM (Cost Per Mille), a standard pricing metric for the cost of $1,000$ ad impressions. The bid floor is an objective variable we use as our price variable to study the price elasticity of supply for the Programmatic Advertising Market.

\item[-] The category(ies) associated with the domain/app generating such a bid request. Categories are extracted from the IAB Content Taxonomy 1.0 \citep{iab_taxonomy}, which is commonly used in programmatic advertising.

\end{itemize}

\subsection{Facebook Advertising Platform}
\label{subsec:backgroud_facebook}

FB is one of the most important closed advertising ecosystems in terms of revenue, accounting for \$84.1B in 2020 \citep{fb_financials}. The \emph{FB advertising platform} operates as a centralized programmatic market, where the supply of ad spaces delivered within the FB ecosystem (i.e., Facebook, Instagram, and Facebook Messenger) \footnote{FB also serves ads to external mobile apps and websites, but this represents a minor portion of its advertising business.} is fully controlled by FB. Advertisers can configure their ad campaigns through the FB Ads Manager using very detailed targeting options related to users' geographical, demographics, and behavioral information. Then, when a user with a profile $\mathit{Pf}$ is going to be exposed to an ad space in any of the FB channels, FB runs an auction \citep{fb_auctions} among those ad campaigns targeting profile $\mathit{Pf}$. Facebook offers different pricing schemes to advertisers, including CPM, which is used as the price reference variable in this paper. \Cref{fig:fb_operation} showcases the operation of the FB advertising platform.

\subsection{Supply-Demand Theory in Online Advertising}
\label{sec:supply-demand-theory}

Supply-Demand theory \citep{marshall2009principles,pindyck2015microeconomics}, is an economic theory that characterizes the behavior of markets based on three parameters: \emph{supply}, \emph{demand}, and \emph{price}.
In online advertising, the goods to be traded are ad spaces. The demand is generated by advertisers willing to buy ad spaces,
while the supply is provided by services from the Open Web (websites, mobile apps, video platforms, social media platforms) which own ad spaces to sell.
Free markets, as the open Programmatic Advertising Market (see Section~\ref{subsec:backgroud_programmatic}), operate to reach an equilibrium at a price $P_{0}$ where supply and demand adjust to each other. Instead, in a monopoly like the FB advertising platform (see Section~\ref{subsec:backgroud_facebook}), a single-player controls the entire supply, and thus it also controls the reaction to a change in demand. Upon a demand change, FB could react by fixing the price and changing the supply, or vice versa. Available information about the FB Advertising Platform operation suggests that FB does not influence the price of ad spaces, which is defined on an auction process based on advertisers' bids \citep{fb_auctions}. However, FB could intervene and deliberately adjust the supply of ad spaces at any moment. 

\begin{figure}[t]
\centering
  \includegraphics[width=0.45\columnwidth]{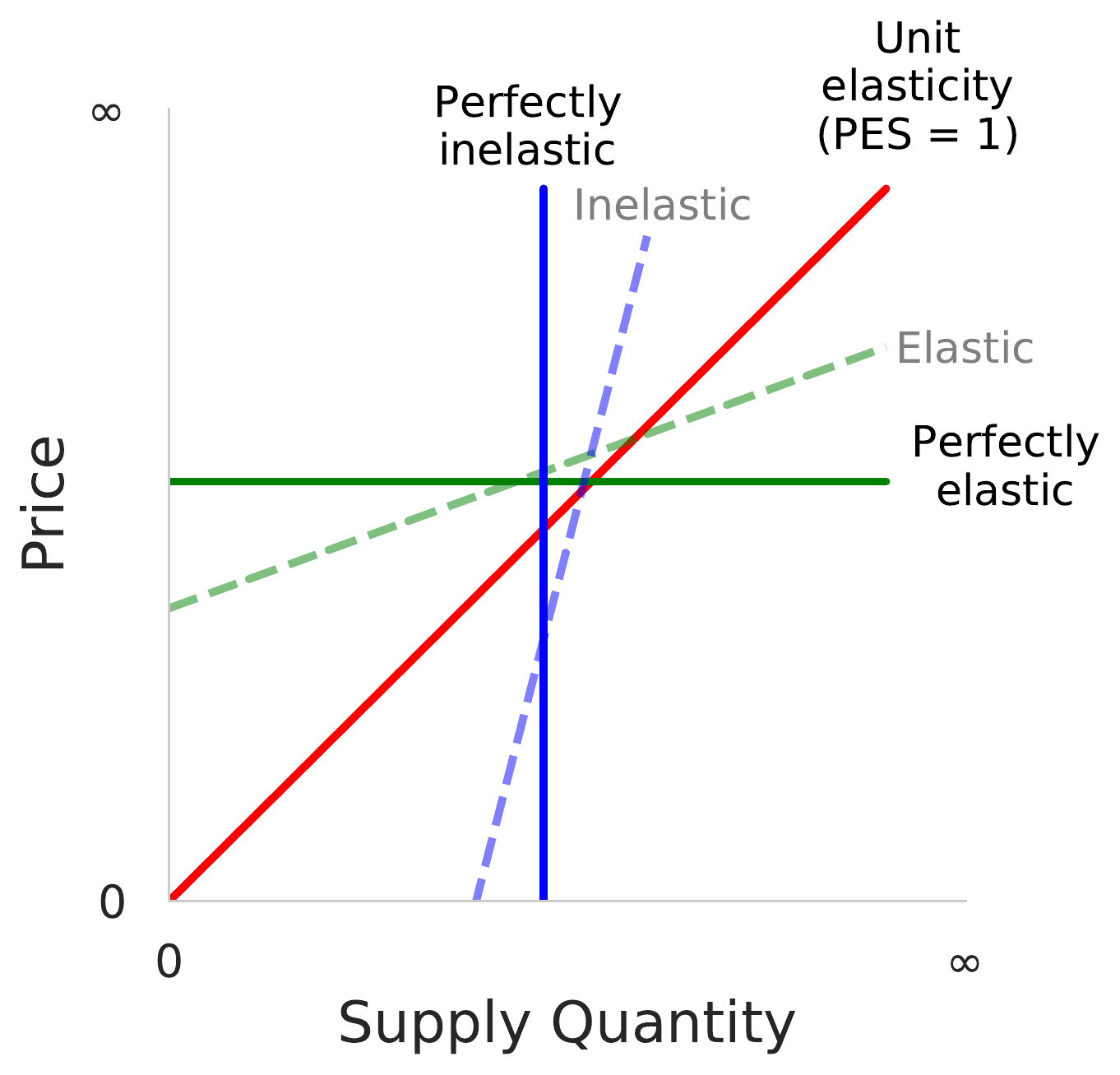}
  \caption{Different supply types based on price elasticity.}
  \label{fig:pes}
\end{figure}

The \emph{elasticity} is a metric that characterizes the sensitivity of the demand or the supply to changes in price. In this paper, we study the elasticity of supply in the online advertising markets. For that, we use the Price Elasticity of Supply (PES).
We determine the PES using the formula of arc elasticity \citep{allen1934concept}, applying the log-log model of Equation~\ref{eq:model} to compute the percentage variation of the supply $S$ as a function of the percentage variation of the price $P$ as expressed in Equation~\ref{eq:pes} ($\beta_1$ refers to the percentage variation between $S$ and $P$). This gives as a result a PES 
in the interval $(-\infty,\infty)$.
\begin{eqnarray}
    \log(supply) \sim \beta_0 + \beta_1\log(price)
\label{eq:model}
\\
    PES = \beta_1 = \frac{\%\Delta supply}{\%\Delta price}
\label{eq:pes}
\end{eqnarray}

The sign of the PES indicates whether supply and price move in the same (positive) or opposite (negative) directions. However, what defines whether the supply is elastic or inelastic is the magnitude of the PES. For that reason, we will only use the absolute value of the PES in the rest of the paper. A PES absolute value in the range $[0,1)$ denotes an inelastic supply, resilient to changes in price. In particular, a PES of $0$ indicates the supply is perfectly inelastic. Contrarily, a PES $\in[1,\infty)$ indicates that the supply is elastic, and thus a change in price affects the supply. The case of PES = 1 and PES = $\infty$ are referred to as unitary elasticity and perfectly elastic supply, respectively. Figure \ref{fig:pes} illustrates the discussed scenarios.

In the context of this paper, if the Open Web presents an inelastic supply during the pandemic, it is an indication of the resilience of the Open Web service to severe perturbations. Contrarily, an elastic supply would suggest that following the drop in prices (and demand) the overall activity of the Open Web service has shrunk. 
\section{Dataset Description}
\label{sec:dataset}

\begin{table}[t!]
\centering
\resizebox{.75\columnwidth}{!}{%
\begin{tabular}{lllll}
\hline
\multicolumn{1}{c}{\textbf{Dataset}} & \multicolumn{1}{c}{\textbf{Size}}                                             & 
\textbf{Area} &
\multicolumn{1}{c}{\textbf{Market}} & \multicolumn{1}{c}{\textbf{Variables}} \\ \hline
Bid requests                        & \begin{tabular}[c]{@{}l@{}}18.2B bid requests \end{tabular}         & Spain & Programmatic                        & Supply, Price                         \\
FB CPM                              & \begin{tabular}[c]{@{}l@{}}14.3B impressions  \end{tabular}     & World & FB advertising                      & Price                                 \\
FB ads                              & \begin{tabular}[c]{@{}l@{}}98.7k ads / 8.8k sessions  \end{tabular} & Spain & FB advertising                      & Supply, Demand                        \\ \hline
\end{tabular}%
}
\caption{Datasets information (Jan.~1st to Dec.~31st, 2020).}
\label{tab:dataset}
\end{table}

As mentioned, we focus our analysis on the Spanish online advertising market. In Spain, the \COVID outbreak arose in early March 2020. The Spanish government established a strict \emph{lockdown} on March 15th, which lasted almost two months, until May 11th. From that date on, a \emph{reopening} plan was implemented by progressively removing lockdown conditions. This plan ended on June 21st, when the state of alarm declared by the government was no longer extended, which led to the called \emph{new normality}. Then, we consider two main phases in our analysis: \emph{preCOVID-19} (between January 1st and March 14th) and \emph{COVID-19} (between March 15th and December 31st). In turn, we split the COVID-19 phase into three sub-phases: \emph{lockdown} (between March 15th and May 10th), \emph{reopening} (between May 11th and June 20th) and \emph{new normal} (between June 21st and December 31st).

In this section, we describe the datasets we use to evaluate the elasticity of both, the Programmatic Advertising Market and the FB Advertising Platform. Moreover, we also introduce the data we use to measure the evolution of the distribution of supply across business categories. Table \ref{tab:dataset} summarizes the most relevant information from these datasets. Finally, at the end of this section we discuss further considerations (see Subsection \ref{sec:considerations_dataset}) and ethical concerns (see Subsection \ref{sec:ethics_considerations}) that may arise from the dataset used in our study.

\subsection{Supply Time Series Data}
\label{sec:supply}

\paragraph{Programmatic Advertising Market}

We use a dataset created from the daily bid request flow received by TAPTAP Digital between January 1st and December 31st, 2020. TAPTAP Digital is a mid-size DSP with a strong presence in the Spanish Programmatic Advertising Market.
This dataset includes an average number of 49.7M bid requests per day obtained from more than 1.5M publishers, including mobile apps and websites, through 9 different AdXs. 
Although we do not know the details of the proprietary algorithm each AdX implements to select which bid requests are sent to a particular DSP, our analysis of the data does not reveal any signal of determinism. Hence, we assume our dataset is a representative sample of the supply of ad spaces in the Spanish Programmatic Advertising Market.

From all the publishers present in the dataset, we have selected those having data available every day in the considered period, so that we can build their complete time series. We are hence left with $2,148$ publishers, which are responsible for 83.5\% of the total bid requests in our dataset.  
For each of these publishers, we have computed $S_P(d,i)$ as the fraction of ad spaces publisher $i$ generates at day $d$. The time series of $S_P(d,i)$ represents the evolution of the relative supply of ad spaces for publisher $i$.

\begin{figure}[t!]
\centering
  \includegraphics[width=.75\columnwidth]{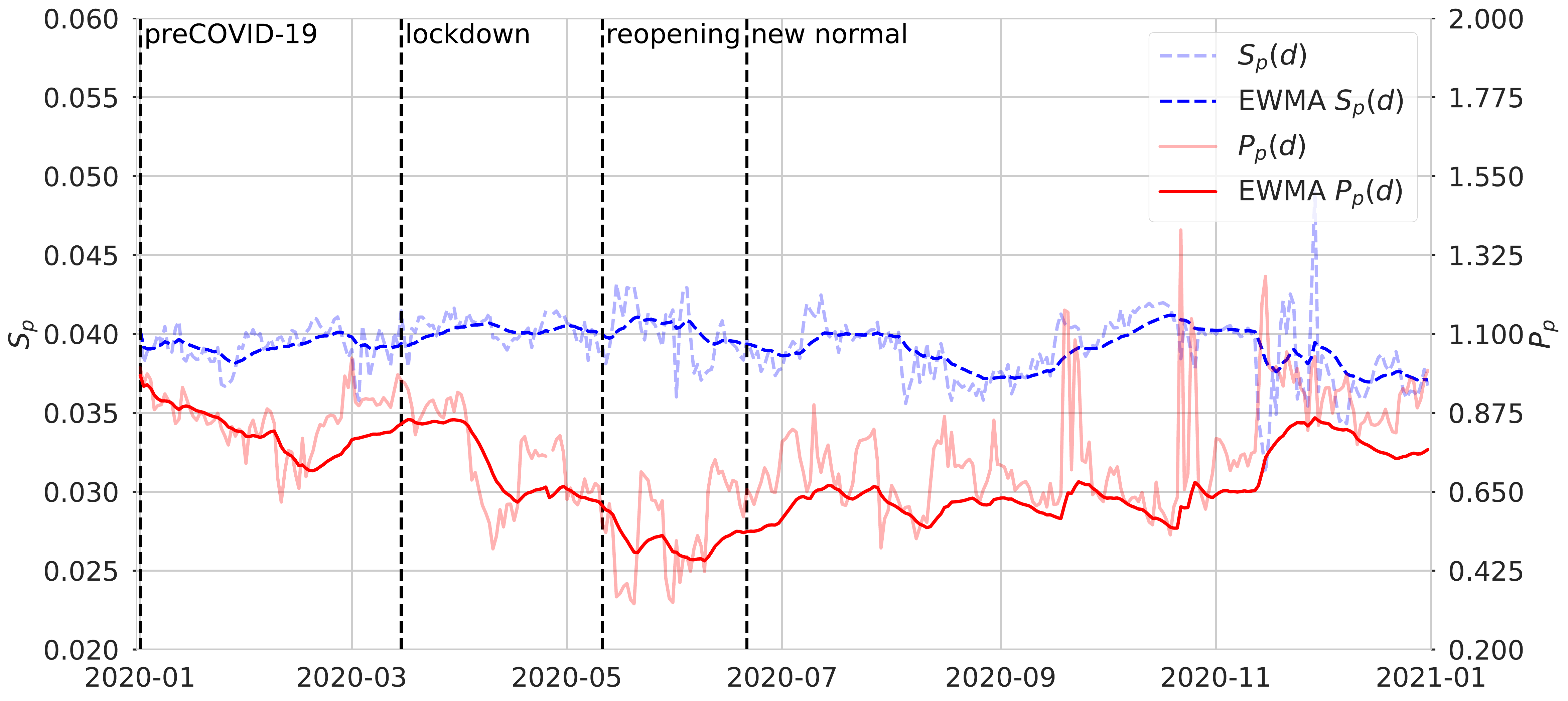}
  \caption{Supply ratio $S_P(d)$ and price $P_P(d)$ (in US\$) time series, and their EWMA for the Programmatic Advertising Market between Jan.~1st and Dec.~31st, 2020.}
  \label{fig:Sp:Pp}
\end{figure}

To evaluate the overall evolution of the supply, we have computed $S_P(d)$ as the average value of $S_P(d,i)$ across the selected publishers. Figure \ref{fig:Sp:Pp} shows the time series of $S_P(d)$ (blue line) for the considered period $ d\in [\mathrm{January\ 1st}, \mathrm{September\ 30th}]$. The figure also presents the Exponential Weighted Moving Average (EWMA) of $S_P(d)$, which captures the trend of the time series.

\begin{figure}[t!]
\centering
  \includegraphics[width=.75\columnwidth]{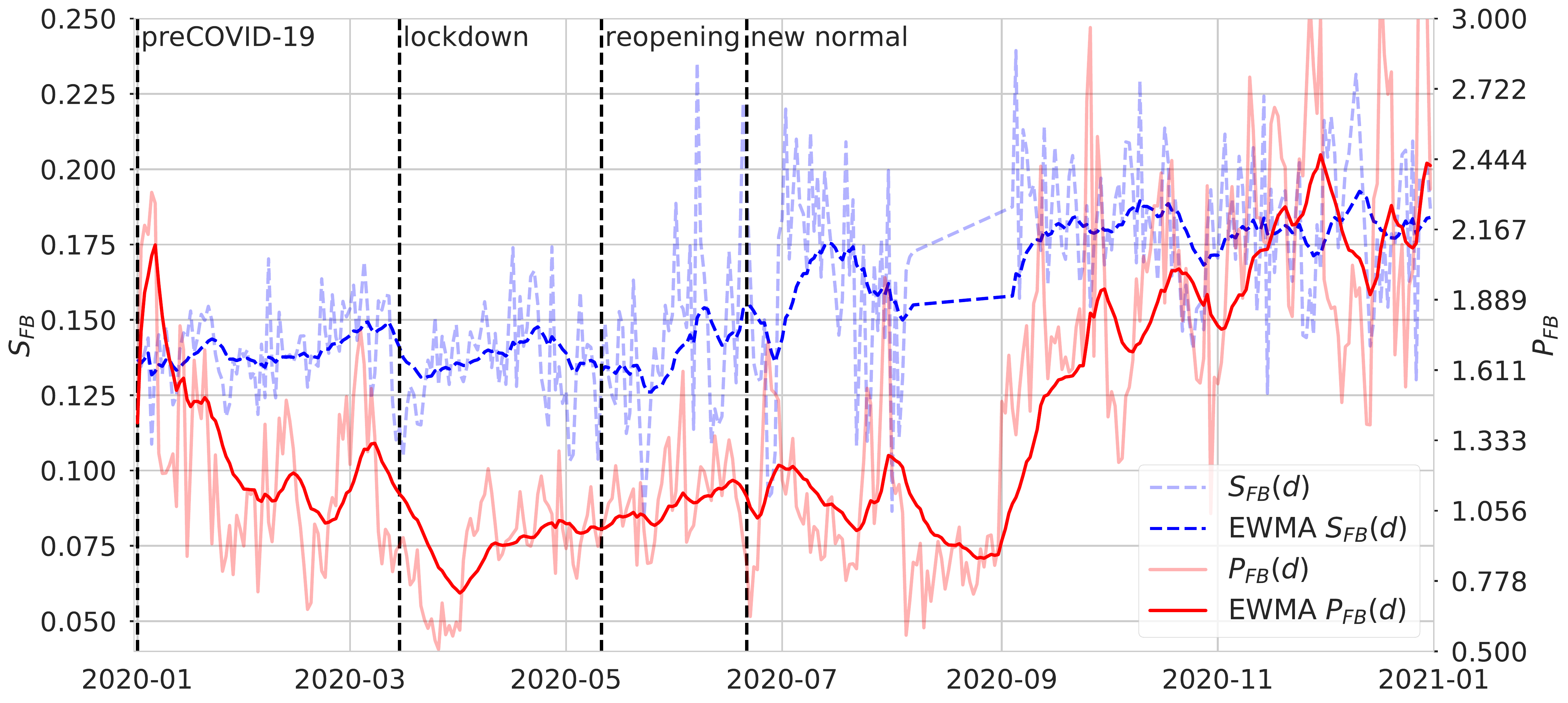}
  \caption{Supply ratio $S_{FB}(d)$ and price $P_{FB}(d)$ (in US\$) time series, and their EWMA for the Facebook Advertising Platform between Jan.~1st and Dec.~31st, 2020.}
  \label{fig:Sfb:Pfb}
\end{figure}

\paragraph{Facebook Advertising Platform} We use data collected from the Data Valuation Tool for Facebook Users (FDVT) browser add-on \citep{FDVT} with more than 9.9k installations worldwide. More specifically, 3k+ users located in Spain have installed this plugin. Our extension was publicly released in Oct. 2016 and our user base is formed by users that freely decided to install it. The main functionality of this browser extension is to provide, with a real-time estimation, the revenue they generate for FB out of the ads they receive while browsing on FB. Before start using the add-on, in the registration process, users provided their country of residence (compulsory). The browser extension also collects (among other data) meta-information (e.g., advertiser domain, timestamp, etc.) of the ads delivered to a user during a FB session and the total number of posts displayed to the user in that session. We use the information of ads and posts delivered to users in 2020 across the sessions to compute $S_{FB}(d,u,s)$, the ratio of ads per information post a given FB user $u$ has been exposed to in a session $s$ during day $d$. For instance, $S_{FB}(d,u,s) = 1/5$ means that user $u$ has received one ad in their newsfeed every $5$ regular posts during session $s$ on day $d$.
$S_{FB}(d)$ is the daily average value of $S_{FB}(d,u,s)$ across a large number of users and sessions. It captures the daily relative supply of ad spaces in the FB Advertising Platform. 
Figure \ref{fig:Sfb:Pfb} shows the evolution of the daily relative supply, $S_{FB}(d)$ (blue line), derived from more than 8.8k sessions of Spanish users running the FDVT browser add-on in the analyzed period, $d \in [\mathrm{January\ 1st}, \mathrm{December\ 31st}]$.\footnote{\label{data-downtime}The FDVT add-on, used to collect Facebook data, did not collect information from August 8th to September 3rd due to a major upgrade that modified the Facebook wall format and operation. It took some time to update the browser add-on to be operative again under the new FB wall format. However, the new normal period (where the data-breach is included) contains a representative amount of more than five-month data for its analysis.} We also present the EWMA of the time series to capture the trend of $S_{FB}(d)$.  

\subsection{Price Time Series Data}
\label{sec:price}

\paragraph{Programmatic Advertising Market} As described in Section \ref{subsec:backgroud_programmatic}, we consider the bid floor (in US\$) as the price variable $P$ in our analysis of the PES for the Programmatic Advertising Market. We have extracted the bid floor information from our bid requests dataset for the same subset of publishers considered in Section \ref{sec:supply}. Altogether, these publishers are responsible (on average) for 34.2M daily bid requests, which include the bid floor information. Figure~\ref{fig:Sp:Pp} shows the time series of the daily average bid floor $P_P(d)$ (red line), and its EWMA, in the considered time window $d \in [\mathrm{January\ 1st}, \mathrm{December\ 31st}]$.

\paragraph{Facebook Advertising Platform} Gupta Media \citep{guptamedia} offers the average CPM value of ad spaces in the FB advertising platform extracted from a large number of FB ads campaigns and provides this data per country since March 2018. We have retrieved the time series of the daily average CPM of ad spaces for Spain in the time window $d \in [\mathrm{January\ 1st}, \mathrm{December\ 31st}]$. The dataset for this period includes information from more than 14.3B ad impressions and 237 countries and regions. The average CPM represents our price variable, $P_{FB}(d)$, for the FB Advertising Platform. Figure \ref{fig:Sfb:Pfb} shows the time series of $P_{FB}(d)$ (see red line) and its EWMA.

\subsection{Supply Distribution across Categories}

As described in Section \ref{subsec:backgroud_programmatic}, a bid request in the Programmatic Advertising Market includes information about the category(ies) associated with the domain/app generating the bid request. Extracting this information from TAPTAP Digital's bid requests flow, we compute the distribution of the (average) daily fraction of ad spaces (i.e., the supply) across categories for each considered phase (preCOVID-19, lockdown, reopening, and new normal). 
Figure \ref{fig:categories:programmatic} depicts the specific composition of the supply across categories for each phase. In particular, it shows the average daily fraction of ad spaces associated with the top 23 categories in each of the four considered phases.

Note that in the case of Facebook's advertising market there is only one supplier of ad spaces, FB. Therefore, it does not make sense to analyze the distribution of ad spaces across supply categories.

\subsection{Considerations Regarding the Datasets}
\label{sec:considerations_dataset}
In this subsection, we discuss in detail a few aspects, which may raise concerns on the validity of our datasets to address the analysis of the supply resilience in online advertising.

\begin{figure}
\centering
\includegraphics[width=.75\columnwidth]{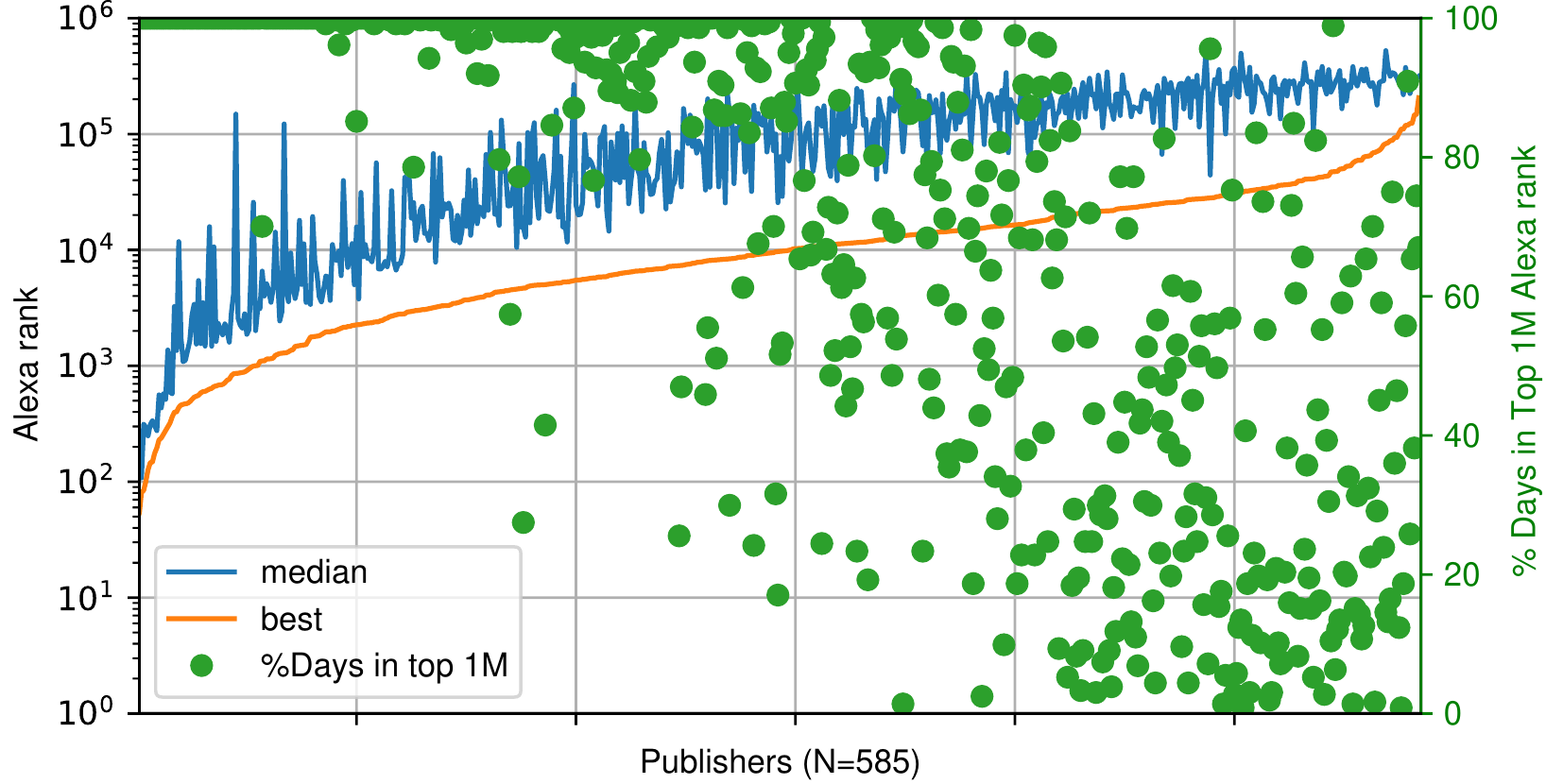}
\caption{Best (orange), median (blue) Alexa rank and the percentage of days (green) that each publisher got indexed in the Top-1M Alexa rank between 1st January and 31st December.}
\label{fig:alexa_best_median_rank}
\end{figure}

\paragraph{Selected publishers representativeness}
We compiled all daily Alexa top sites rankings between January 1st and December 31st
to avoid any bias generated by the Alexa ranking~\citep{scheitle2018long}. 
In Figure~\ref{fig:alexa_best_median_rank}, we present the best, median, and the percentage of days that our publishers got indexed in the top-1M Alexa ranking.
Since Alexa only indexes websites (not mobile apps), this analysis only considers websites from the selected group of 
publishers (585 different websites). We observed that 176 (30\% of the selected publishers) were listed in the top-1M Alexa rank during the whole period covered in this paper; and 30\% of the were always within the top-10k Alexa rank according to the median rank value. These results confirm that the selected publishers span over a wide spectrum of popularity, including 
popular websites like \texttt{soundcloud.com} or \texttt{msn.com} among others.

\paragraph{Bid floor representativeness of market price of ad spaces}

To prove the soundness of the bid floor as a proxy of the market price of ad spaces in the Programmatic Advertising Market, we have measured the correlation between the actual price TAPTAP Digital paid for the auctions they won in the sub-sample between June and September 2020 (extracted from the winning notice message of those auctions) and the bid floor of the corresponding bid request. We obtain a Pearson correlation value of $0.58$ that confirms the bid floor is a reasonably good proxy of the market price of ad spaces. Note that we discarded using only data from the auctions won by TAPTAP Digital (to extract the market value of ad spaces) because they represent a very small fraction of the total collection of bid requests, and may not be representative enough.

\paragraph{Seasonality impact on the market price of ad spaces}

We claim that the observed variability in the price of ad spaces in 2020 is mainly due to the \COVID outbreak. However, one may wonder if such price variability is a seasonal effect happening every year. If this is not the case, and the observed price variation is unique for 2020, we can assume that it is very likely caused by the \COVID outbreak. To test this hypothesis, we analyze the seasonality of the FB price time series\footnote{We do not have historical data for the Programmatic Advertising Market. However, since both FB and the Programmatic Advertising Market are two complementary online advertising channels, we conjecture that the presence/lack of seasonality in FB can be extrapolated to the Programmatic Advertising Market.} between March 2018 and December 2020, comprising a period of more than two years. We have computed the Auto Correlation Function (ACF) of two different periods of the time series: 2020 vs 2019 and 2019 vs 2018. The ACF analysis indicates a lack of significant autocorrelation spikes in the time series with a CI of 95\%. This is translated into the absence of seasonality. We have also checked the seasonal decomposition of the time series with $[30, 60]$ day frequencies, removing the trend component and computing again the ACFs. This leads to even smaller autocorrelation spikes. In a nutshell, the seasonality analysis allows asserting with high certainty that the price fluctuation observed during the \COVID outbreak is most likely due to the pandemic.

\subsection{Ethical Considerations} 
\label{sec:ethics_considerations}
\paragraph{Bid Requests data} We only process data related to domain information from bid requests, and thus we neither use personally identifiable information (PII) nor process any user identifier information. Moreover, the use of the bid requests information is compliant with the terms of use of TAPTAP Digital's providers.

\paragraph{FDVT browser add-on data} The users that have installed the FDVT provided explicit opt-in consent to use the data associated with the ads they receive on FB for research purposes and accepted the Privacy Policy \citep{fdvt_policy_privacy} and Terms of Use \citep{fdvt_terms_of_use}. Also, it is important to note that the add-on used preserves the privacy of users since no PII information is collected.

\section{Open Web Resilience: Programmatic Advertising Market}

In this section, we present the results of our resilience analysis of the Programmatic Advertising Market. First, we obtain the PES of the Programmatic Advertising Market to check if the service offered remains stable during the \COVID pandemic. Then, we study the evolution of the composition of the supply in this market across business categories, to understand whether the distribution among categories has changed during the \COVID pandemic. 

\subsection{Programmatic Advertising Market Elasticity}
\label{sec:elasticity}

Before looking at the Programmatic Advertising Market elasticity, let us briefly evaluate the drop in income this market has suffered.
This income reduction is confirmed by online advertising stakeholders' reports \citep{iab:sellside:COVID,iab:buyside:COVID, pixalate:programmatic:COVID}, but also by the long-lasting low-price situation depicted by the price time series in Figure \ref{fig:Sp:Pp}:

\begin{itemize}
\item[-]The price drop started two weeks after the lockdown was enacted in Spain (March 15th). This represents a quick, but not immediate, reaction to an extreme decision like the severe lockdown imposed in Spain. Our discussion with industry stakeholders indicates that the reason for this two-week delay may be linked to the fact that the programmatic ecosystem's operation is subject to contracts between advertisers, their agencies, and DSPs, which may not allow immediate cancellation of all running ad campaigns.

\item[-]The Programmatic Advertising Market experienced a severe price drop of around $40\%$ between April 1st and May 21st.\footnote{The reported price drop is obtained from the EWMA time series} Since May 21st the Programmatic Advertising Market price has shown a very weak recovery. Indeed, it is not until December 2020, when the price recovered to similar preCOVID-19 values. 
\end{itemize}

Now, let us look at the elasticity of the supply, given this reduction of income in the Programmatic Advertising Market.
We have computed the PES for the Programmatic Advertising Market using the elasticity formula in Equation \ref{eq:pes}. The log-log model of Equation \ref{eq:model} receives as input the logarithmic values for the supply ($S_P$) and price ($P_P$) variables represented in Figure \ref{fig:Sp:Pp}. The PES values obtained are shown in Table \ref{tab:PES_less_publishers} for each of the pandemic periods studied. To understand the elasticity of the online advertising market, we check the magnitude of the PES which, as stated in Section \ref{sec:supply-demand-theory}, is interpreted in absolute terms. The PES value for the Programmatic Advertising Market across the whole pandemic period is $-0.06$. These results indicate that this market, mainly supported by the supply of ad spaces from Open Web players, shows a perfectly inelastic supply in the analyzed period. If we unbundle the analysis of PES for the periods considered in this paper, we found that every period shows an elasticity lower than $0.1$ (in absolute terms). This suggests that the Open Web players have kept their normal operation in the Programmatic Advertising Market during the considered COVID-19 pandemic period, despite the seemingly important reduction in the income. 

\begin{table}[t!]
\centering
\resizebox{.5\hsize}{!}{%
\begin{tabular}{@{}lcccc@{}}
\toprule
\multicolumn{1}{c}{\multirow{2}{*}{\textbf{}}} & \multicolumn{2}{c}{\textbf{FB}} & \multicolumn{2}{c}{\textbf{Programmatic}} \\ \cmidrule(l){2-5} 
\multicolumn{1}{c}{}                           & Elasticity       & 95\% CI      & Elasticity          & 95\% CI         \\ \midrule
preCOVID-19                                    & 0.03             & ±0.07        & -0.05               & ±0.07           \\
Lockdown                                       & 0.06             & ±0.14        & 0.03                & ±0.03           \\
Reopening                                      & 0.22             & ±0.48        & -0.08               & ±0.05           \\
New normal                                     & 0.07             & ±0.07        & -0.09                & ±0.04           \\
All periods                                    & 0.21             & ±0.05        & -0.06              & ±0.02           \\ \bottomrule
\end{tabular}%
}
\caption{Percentage variation of the supply as a function of the percentage variation of the price derived from the log-log model, and its 95\% Confidence Interval (CI), for both the Programmatic Advertising Market and Facebook Advertising Platform. The PES value must be interpreted in absolute values of the elasticity shown in this table.}
\label{tab:PES_less_publishers}
\end{table}

\begin{table}[t!]
\centering
\resizebox{.5\hsize}{!}{%
\scriptsize
\begin{tabular}{@{}lccc@{}}
\toprule
\textbf{} & \textbf{Lockdown} & \textbf{Reopening} & \textbf{New normal} \\ \midrule
Supply    & 0.78              & 0.84               & 0.78                \\
Demand    & 0.65              & 0.64               & 0.76                \\ \bottomrule
\end{tabular}%
}
\caption{Result of the Ružička index across categories for supply and demand when comparing the preCOVID-19 online market status with the next periods of the pandemic.}
\label{tab:jaccards}
\end{table}

\subsection{Distribution of Supply across Business Categories}
\label{subsec:supply_composition}

We now analyze the evolution of the distribution of the business categories in the Programmatic Advertising Market, through their associated supply of ad spaces, during the COVID-19 pandemic.
In particular, we use the distribution of the (average) daily fraction of ad spaces associated with business categories as our reference metric. 
Table \ref{tab:jaccards} presents the Ružička index \citep{ruzicka18anwendung} for this metric between the preCOVID-19 phase and the lockdown (0.78), reopening (0.84) and new normal (0.78) phases, respectively. Note that, the closer the Ružička index is to 1, the more similar are the two compared distributions. 
The results indicate a decrease of the Ružička index. The Ružička index is higher in the most promising period (reopening) as an indication of the thoughts at that time to slowly return to preCOVID-19 lifestyle. However, the New normal index result is a clear indication that the representativeness of business categories and their associated services has been modified by the pandemic. Indeed, at the time of conducting this study, such transformation seems to be still ongoing. 

\begin{figure}[t!]
\centering
     \begin{minipage}{0.485\linewidth}
      \includegraphics[width=\columnwidth]{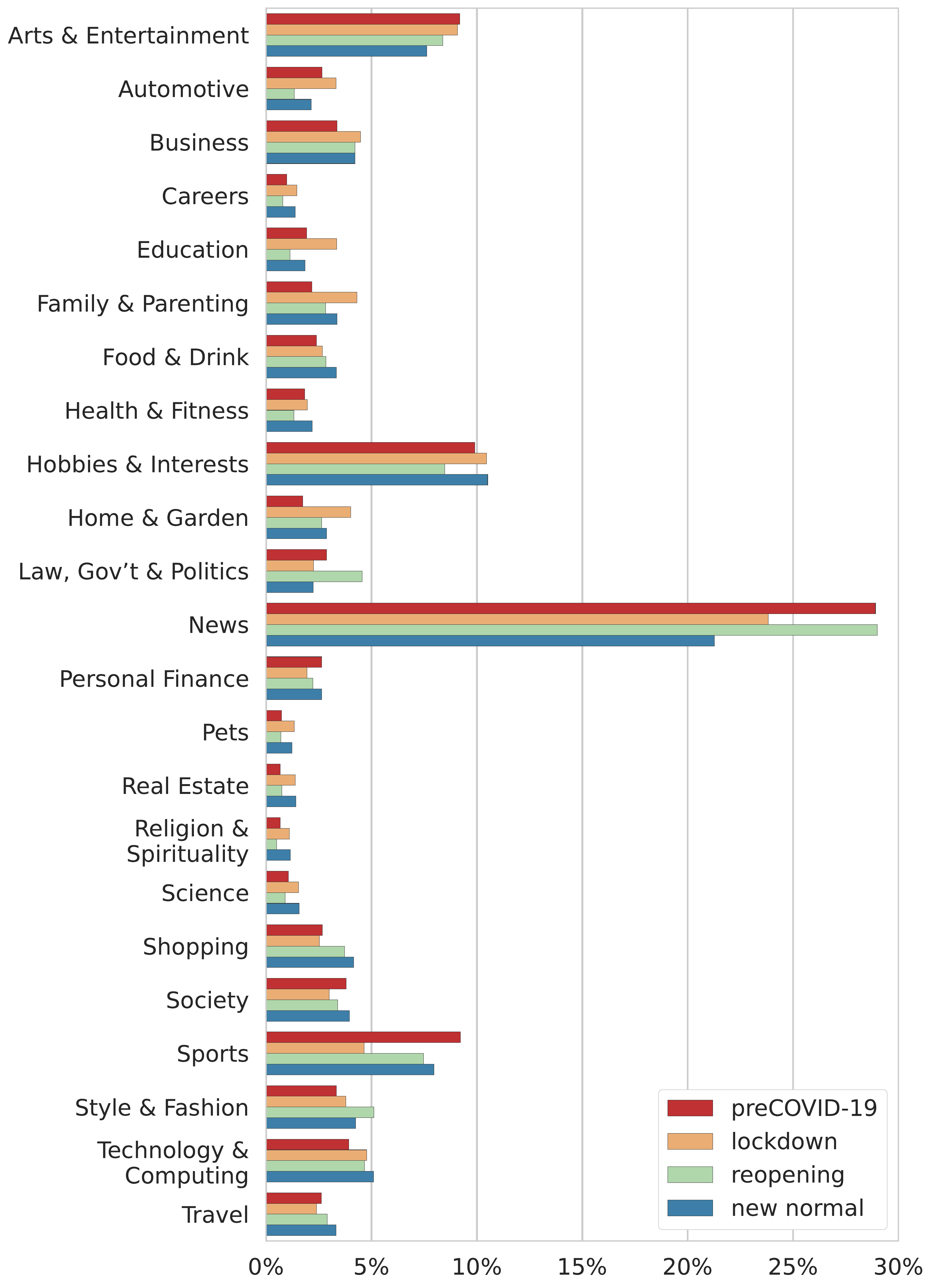} 
      \caption{Distribution of the supply of ad spaces across categories during the preCOVID-19 (red), lockdown (orange), reopening (green), and new normal (blue) phases.}
      \label{fig:categories:programmatic}
  \end{minipage}\quad
  \begin{minipage}{0.485\linewidth}
      \includegraphics[width=\columnwidth]{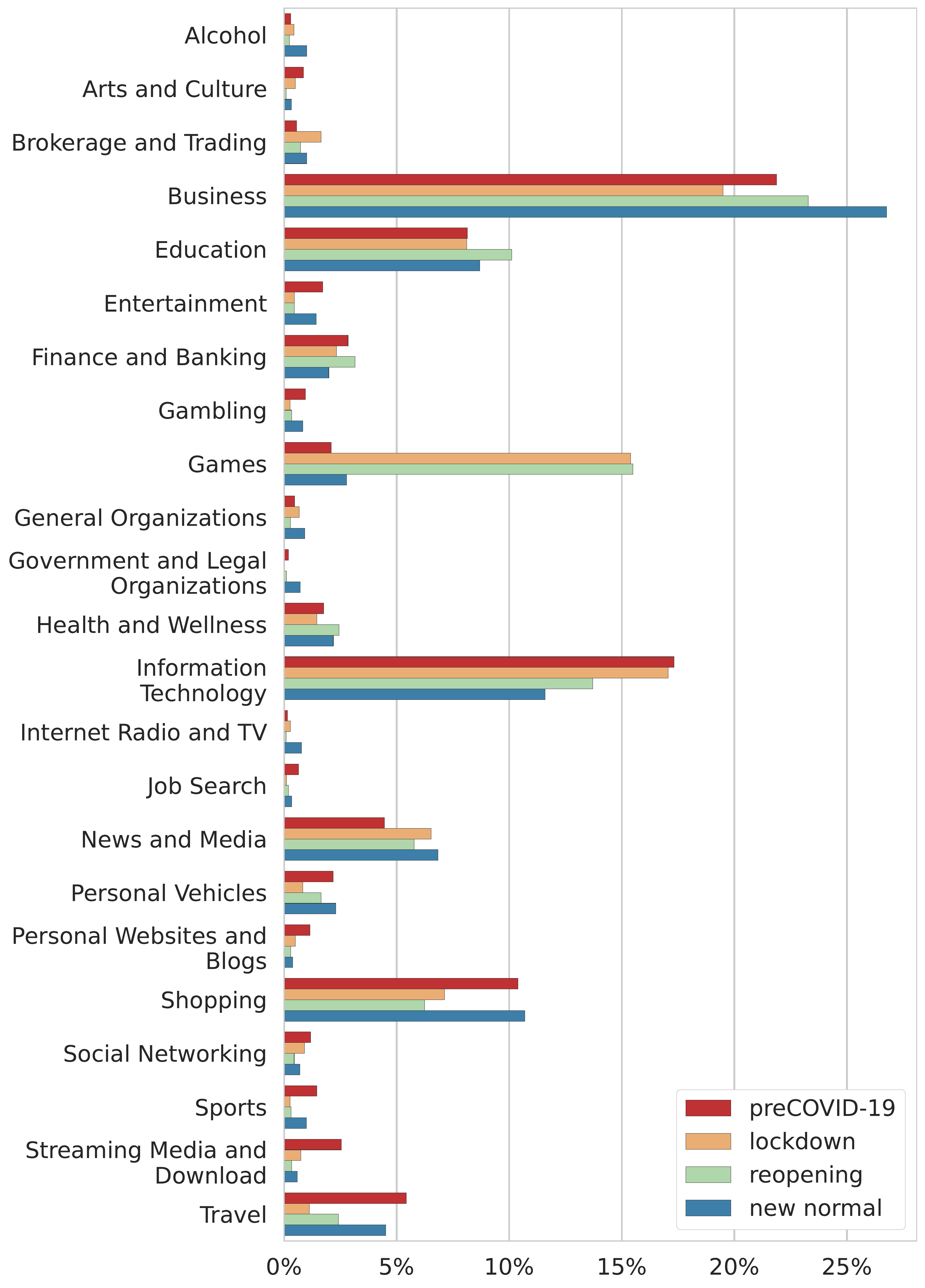}
      \caption{Distribution of the demand of ad spaces across categories during the preCOVID-19 (red), lockdown (orange), reopening (green), and new normal (blue) phases.}
      \label{fig:categories:fb}
  \end{minipage}
\end{figure}

To provide a more detailed view of the evolution of the representativeness of individual categories, Figure \ref{fig:categories:programmatic} shows the distribution of ad spaces associated with each of the top 23 IAB categories for each of the considered phases: preCOVID-19, lockdown, reopening, and new normal. 

We observe a common pattern, in which most categories show an increase (or decrease) in their contribution to the overall supply during lockdown compared to preCOVID-19, which is later reverted in the reopening phase. 
For instance, the \emph{Sports} category contribution drops from $9.2\%$ in the preCOVID-19 phase to just $4.6\%$ in the lockdown phase, and it grows again to $7.5\%$ and $7.8\%$ in the reopening and new normal phases.

It is worth noting that the Ružička index is an objective metric that captures the aggregated evolution of the supply composition. However, individual categories may present their own trends that respond to the specificities of such categories. Discussing the evolution of each category in detail is out of the scope of this paper, but we believe reporting it may be of great value for stakeholders and researchers.

\section{Open Web Resilience: Facebook Advertising Platform}

FB is a dominant player in both the Open Web and the online advertising ecosystems. The Facebook Advertising Platform can be defined as a monopoly since FB has full control of the supply in the platform. Hence, it is interesting to study how the COVID-19 pandemic affected this special player within the online advertising ecosystem. To this end, in this section, we analyze the price elasticity of the supply on Facebook to understand its resilience to the demand and (subsequent) price changes. 
As indicated above, in the case of the FB advertising market it does not make sense to analyze the composition of the supply, since FB is the only supplier of ad spaces. However, our FB dataset allows us to perform an analysis of the composition of the demand across business categories during the pandemic outbreak. This is, how the distribution of ads delivered per business category evolves during the COVID-19 pandemic. 

\subsection{FB Advertising Platform Elasticity}
\label{subsec:elasticity_FB}

Before looking at the elasticity of the FB advertising platform, let us consider its price evolution. As the price time series in Figure \ref{fig:Sfb:Pfb} shows, the impact of the pandemic on FB ads prices is less marked than in the rest of the Open Web:
\begin{itemize}
\item[-]The price of the Facebook advertising market started to fall a few days before Spain established the lockdown (on March 15th). Days before the lockdown was implemented, the Spanish government was taking clear steps towards it (e.g., ceasing all face-to-face academic activity from March 10th), following Italy's example. It seems the FB advertising market reacted based on this information, even before the official lockdown declaration. Note that contrary to the case of the Programmatic Advertising Market, FB operates independently through its Ads Manager that allows advertisers to interrupt their ad campaigns at any moment, which could lead to an immediate reaction to the lockdown.  
\item[-]The FB ads platform experienced a $40\%$ drop in price between March 8th and April 1st. The drop amount is similar to the one reported in the Programmatic Advertising Market. However, after reaching bottom on April 1st, the prices in FB have kept a rather stable growth between April and September (except for August where a new price reduction occurred). Finally, the last quarter of 2020 has brought the ad spaces' price to even higher preCOVID-19 levels. This aligns to the revenue report of FB in 2020. \citep{fb_financials}
\end{itemize}

We computed the PES for the FB Advertising Platform using the same method as for the programmatic market (See Section \ref{sec:elasticity}), but using as input the logarithmic value for the supply ($S_{FB}$) and price ($P_{FB}$) variables represented in Figure \ref{fig:Sfb:Pfb}. The values that resulted from this analysis are shown in Table \ref{tab:PES_less_publishers} for each of the pandemic periods studied. As in the case of the Open Web, we check the PES value and interpret it in absolute terms. The PES values for the FB platform across the whole pandemic period is $0.21$ and $\leq 0.22$ for any of the individual phases. Therefore, the supply of FB presents also an inelastic supply, indicating that, like the rest of the Open Web, its operation is resilient to changes in the demand and price of ad spaces. 

\subsection{Distribution of Demand across Business Categories}

The main goal of this paper is to understand the resilience of the Open Web ecosystem through the analysis of the supply. 
However, as a secondary, and very interesting contribution, we would like to understand how the demand distribution has evolved through the pandemic across different advertising sectors. 
We use our FB dataset for this purpose.
We have classified the landing pages (7.8k fully qualified domain names, FQDN) associated with each of the 98.7k ads collected with the FDVT browser add-on (between January 1st and December 31st) 
into categories, using the FortiGuard Web Filter service \citep{fortiguard}. 
We decided to use FortiGuard for three main reasons highlighted by a recent study on domain classification services \citep{Vallina2020domainclassification}; (1) its accessibility, (2) its output consists of a single label, which eases the analysis, and finally, (3) its wide coverage. After applying this classification 81\% of the 7.8k FQDN have a meaningful label.
Although we acknowledge this data is not a representative sample of the overall online advertising demand in FB, because it only represents the group of users using the FDVT browser add-on, it is still valid to reveal reasonable patterns of demand change across categories.

Replicating the analysis conducted in Section \ref{subsec:supply_composition} for the supply, in this subsection, we have computed the daily average fraction of delivered ads associated with different businesses (i.e., advertising) categories for each of the four phases analyzed in the paper. Using the Ružička index, we have computed the similarity of the distribution of business categories for the preCOVID-19 phase and their counterpart in the lockdown, reopening, and new normal phases. Table \ref{tab:jaccards} presents the Ružička index between the preCOVID-19 phase and the lockdown (0.65), reopening (0.64) and new normal (0.76) phases, respectively. We observe a significant change in the demand composition in the lockdown and reopening phases. After that, it seems the demand is slightly moving back to its preCOVID-19 composition but it is still far from a full recovery.

Finally, Figure \ref{fig:categories:fb} shows the fraction of delivered ads associated with each of the top 23 business categories for each of the four considered phases. 

We observe an important shift in the demand contribution among categories between the preCOVID-19 and COVID-19 phases. As illustrative examples, the \emph{Travel} category contribution drops from $5.4\%$ in the preCOVID-19 phase to only $1.1\%$ in the lockdown and starts growing again in the reopening ($2.4\%$) and new normal phases ($4.5\%$), while the \emph{Games} category faces a notable increase from $2\%$ to $15.4\%$ in lockdown and $15.5\%$ in the reopening phase coming back to $2.7\%$ in the new normal phase. This is aligned with the fact that traveling was not allowed in Spain in the lockdown establishment, but a large fraction of the population spent most of the time at home increasing the online gaming activity. Thus, travel advertisers have incentives to reduce their investment in online advertising while online gaming companies have a clear motivation to increase it during the severe COVID-19 phases.

As in the case of the supply, the evolution pattern of an individual category may differ from the aggregated one defined by the Ružička index. Analyzing the evolution of each category is out of the scope of this paper. However, we make these results available because they may be of value to other researchers as well as stakeholders from the advertising sector.

\section{Related Work}

In the due to understand every angle of this pandemic, the number of research studies about \COVID has rapidly increased from all completely different perspectives. 
The Computer Science discipline has contributed to this effort. For instance, some works have studied the impact the pandemic has had on the traffic and performance of the Internet. To name but a few, Candela~\etal analyzed the increase of the Internet latency in different European countries~\citep{candela2020impact} due to lockdowns. Similarly, Felman~\etal{}, studied the variation in the traffic demand of residential users and the robustness of the infrastructure deployed to respond to the dramatic changes in the demand
~\citep{feldmann2020lockdown}. Boettger\etal investigated the changes in traffic demand from the Facebook perspective, observing for each region differences in performance~\citep{internetReactedCovidFB}.

Other works have focused on studying mobility changes due to the \COVID measures implemented in different countries. Lutu~\etal focused on the user's mobility changes from a mobile network operator perspective in UK~\citep{lutu2020characterization}. Similarly, Schlosser~\etal inspected the mobility changes in German networks using mobility flows collected from
mobile phone data \citep{schlosser2020covid}. Moreover, a whole picture of the European mobility was analyzed by Santamaria~\etal \citep{santamaria2020measuring}. 
Besides, Kuchler~\etal used the Facebook Data for Good dataset, publicly available to the research community to fight against \COVID, to correlate the online and physical iteration in different countries to understand the spread of the disease \citep{kuchler2020geographic}.

Finally, Habes~\etal explored the influence of online advertising to spread healthcare awareness for the \COVID pandemic and analyzed its effectiveness through online surveys~\citep{habes2020relation}.
However, and to the best of our knowledge, the impact the \COVID pandemic has had on the Open Web's and the online advertising business, has not been analyzed so far. 
\section{Discussion and Conclusions}
\label{sec:conclusion}

In this paper, we present a novel analysis of the Open Web response to the \COVID pandemic from the perspective of its financial backbone, the online advertising business.

Our analyses conclude that the Open Web --in Spain-- has experienced a moderate transformation in its business category composition during the pandemic. The distribution of ad spaces across different business categories is differing from the preCOVID-19 period. However, our analysis of the Price Elasticity of Supply demonstrates that this transformation has not been forced by a reduction and financial shortage of the digital advertising activity since the supply of ad spaces is (almost perfectly) inelastic to the reported reduction in demand and price of ad spaces. Instead, the reason for this transformation is related to a shift of the users' browsing behavior and use of mobile apps, which seems to have moderately shaped the Open Web as pointed out by recent works \citep{feldmann2020lockdown, candela2020impact, internetReactedCovidFB}.
A plausible explanation for the observed resilience of the Open Web to the reduction in ad prices is that the marginal cost of maintaining the supply of ad spaces is low compared to the overall cost associated with the service itself. For instance, the fixed costs for running a news media website are mostly dedicated to human capital responsible for generating the content (news) and making it available on the web page. The supply of ads is executed through automated processes that consume mostly hardware resources (e.g., CPU, memory, or network bandwidth), which implies a marginal cost compared to human resources and website operating costs.
As an additional contribution, our analysis reveals that the impact of the \COVID in Facebook (a dominant player in the online advertising market) is significantly less severe than for regular players of the Open Web operating in the open Programmatic Advertising Market.  

We plan to extend the current analysis over the next months (or years) to characterize the reaction of the Open Web ecosystem to the full \COVID pandemic in Spain. Besides, we will replicate our study in a heterogeneous set of countries that have been impacted by the pandemic differently, to better understand how a global unprecedented event like the \COVID outbreak has impacted the online advertising market, and subsequently the Open Web, in different countries.

\section*{Acknowledgments}
The research leading to these results has received funding from: the European Union’s Horizon 2020 innovation action programme under the grant agreement No 871370 (PIMCITY project); the Ministerio de Educación, Cultura y Deporte, Spain, through the FPU programme (Grant FPU16/05852); the Ministerio de Economía, Industria y Competitividad, Spain, and the European Social Fund(EU), under the Ramón y Cajal programme (Grant RyC-2015-17732); the Ministerio de Ciencia e Innovación, Spain, under the project ACHILLES (Grant PID2019-104207RB-I00) and under the project ECID (Grant PID2019-109805RB-I00, cofunded by FEDER); the Fundación BBVA under the project AERIS; TAPTAP Digital-UC3M Chair in advanced AI and data science applied to advertising and marketing; the Comunidad de Madrid (CM), Spain, synergic project EMPATIA-CM (Grant Y2018/TCS-5046) and EdgeData-CM (Grant P2018/TCS4499, cofunded by FSE \& FEDER); the IMDEA Networks and Comunidad de Madrid (CM), Spain, project CoronaSurveys-CM; and the Comunidad de Madrid and Universidad Carlos III de Madrid for the funding of research projects on SARS-CoV-2 and COVID-19 disease, project name ``Multi-source and multi-method prediction to support COVID-19 policy decision making'', which was supported with REACT-EU funds from the European regional development fund ``a way of making Europe''.

\bibliographystyle{elsarticle-harv} 
\bibliography{cas-refs}





\end{document}